\documentclass[sigconf,authordraft,review=false]{acmart}

\AtBeginDocument{%
  \providecommand\BibTeX{{%
    \normalfont B\kern-0.5em{\scshape i\kern-0.25em b}\kern-0.8em\TeX}}}

\copyrightyear{2023} 
\acmYear{2023} 
\setcopyright{none}
\acmPrice{15.00}
\usepackage{caption,nccmath}
\usepackage{subcaption}
\usepackage[titlenumbered,ruled]{algorithm2e}
\usepackage{multirow}
\usepackage{arydshln}
\usepackage{amsmath}

\DeclareMathOperator*{\argmin}{arg\,min}
\usepackage{cleveref}

\begin{document}

\title{Causal Decision Transformer
for Recommender Systems via Offline Reinforcement Learning}

\author{Siyu Wang}
\affiliation{%
  \institution{The University of New South Wales}
  \city{Sydney}
  \country{Australia}
  \postcode{2052}
}
\orcid{0009-0008-8726-5277}
\email{siyu.wang5@student.unsw.edu.au}

\author{Xiaocong Chen}
\affiliation{%
  \institution{The University of New South Wales}
  \city{Sydney}
  \country{Australia}
}
\email{xiaocong.chen@unsw.edu.au}

\author{Dietmar Jannach}
\affiliation{%
 \institution{University of Klagenfurt}
 \city{Klagenfurt}
 \country{Austria}}
\email{dietmar.jannach@aau.at}

\author{Lina Yao}
\affiliation{%
  \institution{Data61, CSIRO}
  \city{Eveleigh}
  \country{Australia}
}
\affiliation{%
  \institution{The University of New South Wales}
  \city{Sydney}
  \country{Australia}
}
\email{lina.yao@unsw.edu.au}

\renewcommand{\shortauthors}{Wang, et al.}

\begin{abstract}
  Reinforcement learning-based recommender systems have recently gained popularity. 
 However, the design of the reward function, on which the agent relies to optimize its recommendation policy, is often not straightforward. 
 Exploring the causality underlying users' behavior can take the place of the reward function in guiding the agent to capture the dynamic interests of users.
  Moreover, due to the typical limitations of simulation environments (e.g., data inefficiency), most of the work cannot be broadly applied in large-scale situations. Although some works attempt to convert the offline dataset into a simulator, data inefficiency makes the learning process even slower. Because of the nature of reinforcement learning (i.e., learning by interaction), it cannot collect enough data to train during a single interaction. Furthermore, traditional reinforcement learning algorithms do not have a solid capability like supervised learning methods to learn from offline datasets directly. 
  In this paper, we propose a new model named the causal decision transformer for recommender systems (CDT4Rec). CDT4Rec is an offline reinforcement learning system that can learn from a dataset rather than from online interaction. Moreover, CDT4Rec employs the transformer architecture, which is capable of processing large offline datasets and capturing both short-term and long-term dependencies within the data to estimate the causal relationship between action, state, and reward. To demonstrate the feasibility and superiority of our model, we have conducted experiments on six real-world offline datasets and one online simulator.
\end{abstract}




\maketitle

\section{Introduction}
Reinforcement Learning (RL)-based Recommender Systems (RS) have been proven effective in a wide range of applications, including e-commerce, advertising, and streaming services, especially considering that users' interests are constantly changing in the real world~\cite{chen2021survey}.
In RLRS,
an agent interacts with the environment by taking actions (e.g., recommending items to users) and receiving feedback in the form of rewards (e.g., user actions on recommended items). The agent uses the feedback to improve its policy over time, with the ultimate goal of maximizing the long-term reward (e.g., increasing user satisfaction or engagement with the system). However, RLRS often face two major challenges: i) The reward function used to reflect the user's interest is hard to formulate; 
ii) Due to the nature of the reinforcement learning algorithm,  it can only be trained in small-scale simulation platforms, and most of the large existing datasets can not be directly used to optimize the recommendation algorithms.

In RL, the reward function plays a crucial role in evaluating the effectiveness of the current action. When it comes to RLRS, the reward function is used to assess whether a recommended item is a good fit for a specific user. This evaluation can be considered a representation of the user's interests or behavior logic. However, defining the reward function can be a challenging task. Some researchers have chosen to omit it altogether and instead learn the recommendation policy directly from expert demonstrations~\cite{chen2021generative,chen2022generative}. This approach, however, requires pre-training an expert agent in a simulation environment. Unfortunately, the lack of suitable simulation environments makes achieving this goal challenging.
Meanwhile, the majority of recommendation algorithms proposed in the literature rely on data-driven approaches, with offline datasets used for both training and testing. However, such an approach is incompatible with the reinforcement learning paradigm, which necessitates online training and evaluation. 
To address the aforementioned two challenges, one possible solution is to integrate the established data-driven approach into the reinforcement learning framework and avoid the design of reward function.

Recently, data-driven methods, such as transformers~\cite{vaswani2017attention}, have attracted significant attention. Transformers are renowned for their ability to handle large datasets, and the success of BERT4Rec~\cite{sun2019bert4rec} demonstrated the effectiveness of transformers in recommendation systems. Further research~\cite{zhou2020s3,chen2019behavior} has shown that transformers can effectively handle sparse and high-dimensional data. Therefore, in this work, we investigate the use of transformers in RLRS to enhance their capacity for processing large datasets.
Decision Transformer~\cite{chen2021decision} and Trajectory Transformer~\cite{janner2021offline} are two typical offline reinforcement learning methods that leverage transformers. Both approaches view offline RL as a sequence modeling problem and train transformer models on collected data. 
Decision Transformer shows that by training an autoregressive model on sequences of states, actions, and returns, agents can learn to generate optimal behavior with limited experience~\cite{chen2021decision}. 
On the other hand, the Trajectory Transformer demonstrates that sequence modeling in reinforcement learning is a more reliable method for predicting long-term outcomes in environments that satisfy the Markov property~\cite{janner2021offline}.

However, both Decision Transformer and Trajectory Transformer were designed primarily for robot learning. They may not be suitable for addressing the unique challenges faced by RLRS (i.e., understanding user behavior and learning users' dynamic interests). Addressing these challenges requires understanding the causal logic behind their observed behaviors.
Hence, we develop a causal mechanism in the transformer to estimate the causal relationship between action, state, and reward to predict users' potential feedback on the actions taken by the system. In this way, we can avoid designing a reward function and instead estimate the reward by relying on the causality in the user's recent behavior.

The contributions of this work are as follows:
\begin{itemize}
    \item To avoid the reward function design, we design a causal mechanism to estimate the reward based on the user's recent behavior.
    \item We propose a method named causal decision transformer (CDT4Rec), which adopts transformer and offline reinforcement learning as the main framework to empower the proposed method to use existing real-world datasets.
    \item To the best of our knowledge, this is the first work that uses offline reinforcement learning and transformer in a recommender system.
    \item Extensive experiments on six public datasets and one online environment demonstrate the superiority of the proposed method.
\end{itemize}

\section{Problem Formulation}
Let $\mathcal{U} = \{u_0, u_1,...,u_n\}$ denote a set of users
and $\mathcal{I} = \{i_0, i_1,...,i_m\}$ denote a set of items.
The datasets in a recommendation problem consist of the information and historical interaction trajectories of users
spanning time steps $t = 1,..., T$.

A standard recommendation problem can be described as an agent aiming to achieve a specific goal, 
which learns from interactions with users, such as recommending items and receiving feedback.  
This process can be formulated as a RL problem, where an agent is trained to interact with an environment. 
Normally, RL is described as a Markov Decision Process (MDP)~\cite{sutton2018reinforcement}. 
Specifically, the agent interacts with the environment continually, choosing actions based on the current state of the environment and the environment responding to those actions and providing a new state to the agent.
Formally, components mentioned above in the MDP can be represented as a tuple $(\mathcal{S, A, P, R, \gamma})$, in which:
\begin{itemize}
    \item State $\mathcal{S}$: state space. $s_t \in \mathcal{S}$ is the state in time step $t$, and $s_T$ is the terminal state, where $T$ is the ﬁnal time step of an episode.
    \item Action $\mathcal{A}$: action space. $\mathcal{A}(s_t)$ is set of actions possible in state $s$.
    \item Transition Probability $\mathcal{P}$: denoted as $p(s_{t+1}|s_t, a_t) \in \mathcal{P}$, is the probability of transitioning to state $s_{t+1}$, from $s_t$ with $a_t$.
    \item Reward $\mathcal{R}$: $\mathcal{S} \times \mathcal{A} \to \mathbb{R}$ is the reward distribution, where $R(s, a)$ is the reward that an agent receives for taking action 
    $a$ when observing the state $s$.
    \item Discount-rate Parameter $\mathcal{\gamma}$: $\mathcal{\gamma} \in [0, 1]$ is the discount factor for future rewards.
\end{itemize}
Given $(\mathcal{S, A, P, R, \gamma})$, an RL agent behaves following its policy $\pi$, which is a mapping from states to actions to be taken when in those states. 
The RL objective, $J(\pi)$, can then be written as an expectation under the trajectory distribution $(s_0, a_0,..., s_T, a_T)$:
\begin{equation}
\label{eq:rl}
J(\pi) = \mathbb{E}_{\tau \sim p_\pi(\tau)} \bigg[\sum_{k=0}^{\infty} \gamma^k r(s_t, a_t)\bigg]
\end{equation}

\emph{Offline} reinforcement learning is defined as the data-driven formulation of RL problems that improve with more data training.
In an offline RL problem, the goal is still to train the agent to maximize the total reward it receives over time, as expressed in ~\Cref{eq:rl}. 
The fundamental difference between offline RL and RL is that offline RL just uses offline data and does not require any further online interaction~\cite{levine2020offline}.
Thus, a static dataset of transitions, such as data collected previously or human demonstrations, is presented to the offline RL learning system. As a result, the agent in offline RL is deprived of the ability to explore and interact with the environment to collect more transitions.
Formally, we consider the transition dataset $\mathcal{D}$ 
to be the training dataset for the policy, from which the offline RL algorithms try to gain adequate insight into the dynamical systems underlying the MDP.
We denote the distribution over states and actions in $\mathcal{D}$ as $\pi_{\beta}$. 
The states in state-action pairs $(s,a)\in \mathcal{D}$ are sampled following $s \sim d^{\pi_{\beta}}(s)$, and the actions are sampled following $a \sim \pi_{\beta}(a|s)$.

As a result, a dataset for an offline RL-based recommender system can be described formally as $\mathcal{D} = \{(s_t^u, a_t^u, s_{t+1}^u, r_t^u)\}$, following the MDP $(\mathcal{S, A, P, R, \gamma})$. 
For each user $u$ at the time step $t$, we have the following elements: a current state $s_t^u \in \mathcal{S}$, items recommended by the agent (or recommender system) via taking action $a_t$, and the user's feedback $r_t^u$.

\section{Methodology}
\subsection{Model Architecture}
\begin{figure*}[h]
  \centering
  \includegraphics[width=\linewidth]{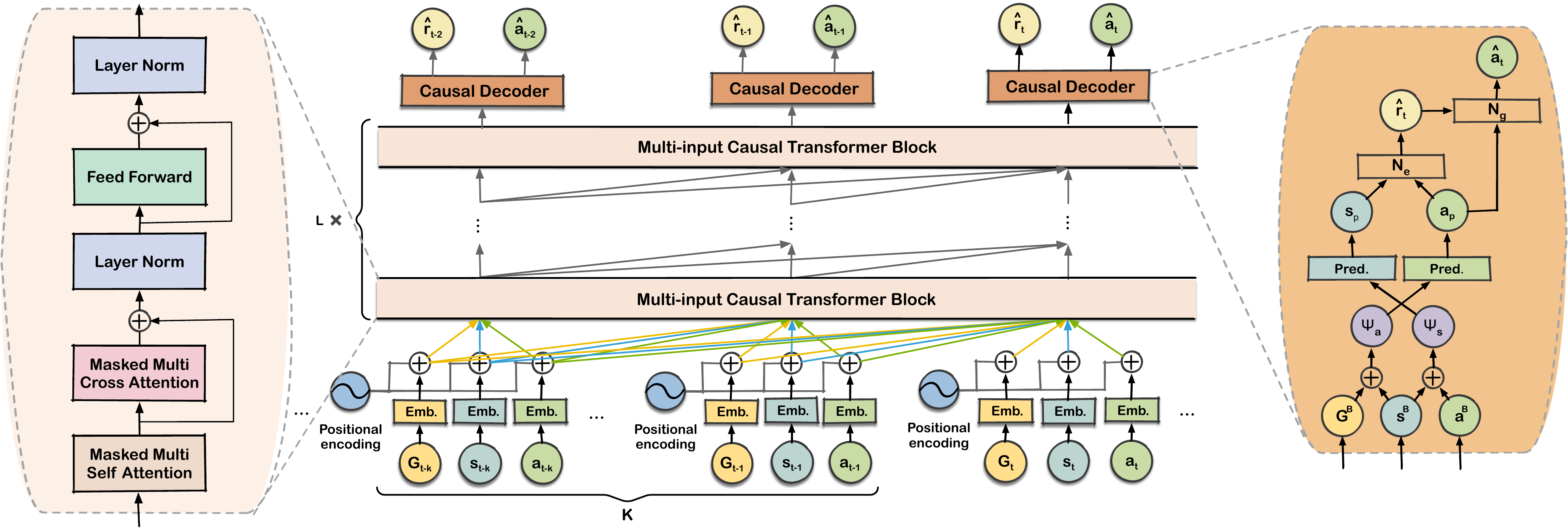}
  \caption{An overview of our CDT4Rec architecture. RTGs, states, and actions are fed into linear embeddings, to which an absolute positional embedding is added. The tokens are then passed through L stacked multi-input blocks, which use both self-attention and cross-attention mechanisms with causal masks. This results in the generation of two balanced representations at the output of the L-th block. These representations are then fed into two additional 
prediction layers. On top of that are two networks $N_e$ and $N_g$, which are used to estimate the reward and generate the predicted action, respectively.}
  \Description{.}
  \label{overrall}
\end{figure*}

Here, we introduce a new 
\textbf{C}asual \textbf{D}ecision \textbf{T}ransformer for RL-based \textbf{R}ecommender \textbf{S}ystem 
(\textbf{CDT4Rec}), which formulates the problem of offline reinforcement learning as a sequence modeling problem. Our framework is built upon Decision Transformer~\cite{chen2021decision}.
As illustrated in ~\Cref{overrall}, CDT4Rec consists of 
$L$ stacked multi-input transformer blocks.
CDT4Rec receives a trajectory representation as the input to the first transformer block.
The output sequence from the L-th transformer block is utilized to construct two distinct final representations: one for action prediction and the other for state prediction.
Additionally, two separate prediction layers are used above the final representations to make predictions for actions and states, respectively.
After that, we train two networks to estimate the reward and generate the action.

For training, we have access to observational data $\mathcal{D}$ and define a trajectory representation as a sequence of three tokens: returns-to-go (RTG), states, and actions. For simplicity, we omit the user index $u$ unless necessary.
Formally, the trajectory representation for autoregressive learning and generation is as follows:
\begin{equation}
\label{eq:tau}
\tau = (G_1, s_1, a_1, G_2, s_2, a_2,..., G_T, s_T, a_T),
\end{equation}
where the RTG $G_t$ is defined as
the sum of the discounted rewards that the agent receives over the future:

\begin{equation}
\label{eq:G}
G_t = r_{t} + \gamma r_{t+1} + ... + \gamma^{T-t} r_{T}  = \sum_{k=t}^{T} \gamma^{k-t} r_k
\end{equation}
The reason for choosing the expected discounted return is that we anticipate the proposed framework to generate future actions during test time that would yield the desired returns.
The generated actions should be compatible with the selection of the agent, which aims to maximize the expected discounted return. 

Let $K \geq 1$ denote the context length, representing the latest $K$ time steps fed into the RL causal transformer. Accordingly, $s_{t-K+1:t}$ represents the sequence of the last $K$ states at time $t$, and $G_{t-K+1:t}$ represents the expected discounted return for the last $K$ time steps. 
Further, let $B \geq 1$ denote the prediction range of the B-step prediction, and $a_{t:t+B-1}$ denote the associated length-B actions generated by the transformer. 
Specifically, 
the generated action at time $t$ is based on $s_{t-K+1:t}$ and $G_{t-K+1:t}$, which can be formulated as 
$a_{t} = \pi(s_{t-K+1:t}, G_{t-K+1:t})$.

In a recommender system, the user's actions are usually represented as binary, either a click or no click. However, this representation of the action is not informative enough to understand the user's interests. Learning to generate actions based on this type of dataset is also inadequate for explaining the user's click behavior.
In RL, on the other hand, the reward function is frequently seen as the most succinct description of a task. The expected reward is used to construct the learning objective for the reinforcement learning algorithm. As a result, the reward function describes the agent's goals and directs policy optimization.
The reward function is critical everywhere, and this is also the case when using RL for RS.
In RLRS, the reward function explains a user's behavior and reflects their interests. In the context of RS, the reward function is designed to partially reflect the logic behind user actions, similar to the concept of causality, which is used to identify the impact of different variables on a model's performance and to understand the reasons behind certain decisions or actions.
In this work, we formulate the reward function as a problem of estimating the causal effects of the action on the reward by giving user history trajectories: $\mathbb{E}(r(a_t, s_t)|\tau)$.

In the following sections, the major components of CDT4Rec are introduced bottom-up: embedding layer, transformer layer, and causal layer.


\subsection{Embedding Layer}
We learn a linear layer to obtain token embeddings for RTGs, states, and actions, respectively, which is followed by layer normalization as in ~\cite{chen2021decision}. 
In addition, 
to determine the exact positions of these tokens within a trajectory,
we apply position encoding to the input embeddings at the bottoms of the transformer layer stacks. 
This is essential for capturing the causality behind the users' behavior as one user's behavior is causally determined by the previous one and the recommendation made by the system.
And the order of users' behaviors can reflect their actual interests.
For example, some users begin their search with higher-priced items but end up purchasing more cost-effective items, while others begin their search with lower-priced items but purchase items from well-known companies. The time step information in the trajectories may reflect different user preferences and habits.

Because the RTG sequence for the recommender system lacks time step information, we must learn an embedding for each time step in order to stitch the sub-trajectories together. 
To help determine the exact positions of tokens inside a trajectory, we use an absolute positional embedding that uses a linear layer to obtain the embedding for each time step in a trajectory. 

Specifically, a one-time-step embedding is appended to three tokens: RTG, state, and reward. And the input representation for each token is constructed by summing its embedding and positional embedding:
\begin{equation}
\begin{gathered}
\label{eq:input1}
G^0_t = \text{Linear}_G(G_t) + p_t\\
s^0_t = \text{Linear}_s(s_t) + p_t \qquad a^0_t = \text{Linear}_a(a_t) + p_t,
\end{gathered}
\end{equation}
where $p_t \in P$ is the $d$-dimensional absolute positional embedding for time step $t$. 

In addition to the absolute positional embedding we used, there is another form of positional embedding called relative positional embedding~\cite{dufter2022position}. 
Relative positional embedding is defined as encoding the position of a unit relative to other units by considering the distance between two tokens. 
\citet{huang2020improve} analyzed the storage complexity of absolute and relative position embedding methods. 
Consider the following transformer model: $b$ layers, $h$ attention heads per layer, and a maximum sequence length of $m$. The absolute and relative position embedding parameter sizes are $md$ and $bh(2m-1)d$, respectively. And the runtime storage complexity is $\mathcal{O}(md)$ for the absolute method and $\mathcal{O}(bhm^2d)$ for the majority of the relative methods. 
Earlier research suggests that relative position embeddings outperform absolute position embedding in some situations~\cite{huang2020improve, melnychuk2022causal}. 
However, given the nature of reinforcement learning and the intricacy of the recommendation problem, adding too much complexity via positional embedding may make RL training harder to converge. 
Moreover, the $m$ we used in this paper is related to the length of users' history and may easily reach very large values, which may easily result in a too-large space complexity for practical problems.
As a result, we employ absolute positional embedding in our method.



\subsection{Transformer Layer}
As demonstrated by~\Cref{overrall}, our transformer layer contains $L$ identical blocks. Let $l = 1,..., L$ be the index of the transformer blocks from bottom to top. For time step $t$, we iteratively compute hidden representations simultaneously at each layer $l$ for each RTG, state, and action, denoted by $G^l_t$, $s^l_t$ and $a^l_t$, respectively. 
The input for each transformer block is three parallel sequences of hidden representations. 
Let $d_h$ be the size of the hidden states and $H^l = (H^l_1, ....H^l_T)^\top \in \mathbf{R}^{T \times d_h}$ be the hidden representations of any one of the $G^l$, $s^l$ and $a^l$ for each layer $l$ at each time step $t$. 
The first block, in particular, is fed with the sequence of ~\Cref{eq:input1} as input. The input for all blocks with $l \geq2$ is the output of the previous $(l-1)$ block:
\begin{equation}
h_t^l= \text{transformer\_block}(h_t^{l-1}),\quad \text{for}~l\geq2
\end{equation}
Following the transformer layer from~\cite{chen2021decision, dong2021attention}, each transformer block begins with a Masked Multi-Head Self-Attention sub-layer over the input tokens followed by a Masked Multi-Head Cross-Attention sub-layer and a Position-wise Feed-Forward layer.

{\bf Masked Multi-Head Self-Attention.} 
Rather than performing a single attention function, we use multi-head self-attention, which is scaled dot-product attention that adapts numerous concurrent attention heads. 
Benefitting from the multi-head attention, the model can pay attention to inputs from many representation subspaces simultaneously at distinct positions.
The attention head takes as input a matrix of queries $Q$, a matrix of keys $K$, and a matrix of values $V$, where $Q, K, V  \in R ^{T\times (d_h/n_h)}$ and $n_h$ is the number of heads. The query, key, and value embeddings are projected from the input tokens with different projection matrices.
Then the attention function is computed on a collection of queries at the same time. The output matrix is as follows:
\begin{equation}
    \text{Attention}^{(j)}\big(Q^{(j)}, K^{(j)}, V^{(j)}\big) = \text{softmax} \Bigg(\frac{Q^{(j)}K^{(j)\top}}{\sqrt{d_{qk}}} \Bigg)V^{(j)}
\end{equation}
The output representation produced by each head is:
\begin{equation}
    \text{head}^{(j)} = \text{attention} \bigg(Q^{(j)}W^Q_j, K^{(j)}W^K_j, V^{(j)}W^V_j\bigg),
\end{equation}
where the projections for each head $W^Q_j, W^Q_j, W^Q_j \in R ^{d_h \times (d_h/n_h)}$ are the learnable parameter matrices.

Multi-head attention, in particular, applies the $h$ attention functions in parallel to build output representations and then outputs the projected concatenation of the different heads:
\begin{equation}
    \text{MultiHead}(Q, K, V) = \text{Concat}(\text{head}^1, ..., \text{head}^h)W^O
\end{equation}
We adopt a variant of multi-head attention where certain input positions are masked with a causal mask~\cite{chen2021decision}.
This is used to ensure that the attention mechanism does not attend to future tokens when generating a predicted action sequence, 
such that autoregressive models can be applied without violating the assumption that future tokens are not available at the time of creation.

{\bf Masked Multi-Head Cross-Attention.} 
In contrast to the self-attention layer, which only takes into account the information within a single transformer subnetwork, the cross-attention layer is capable of exchanging information between parallel transformer subnetworks globally. Specifically, 
self-attention computes all of the keys, queries, and values using hidden states from a single transformer subnetwork for the attention calculation. 
On the other hand, cross-attention infers the queries based on the hidden states within the transformer subnetwork and uses the output from the self-attention layers in two other transformer subnetworks as keys and values. This allows the cross-attention layer to exchange information between all three transformer subnetworks.
To prevent queries from attending to future keys and values, a causal mask is introduced to the cross-attention as well.

{\bf Position-wise Feed-Forward Network.} 
To transform the output of the cross-attention layer into a new representation that is more suitable for the task, we employ a Position-wise Feed-Forward Network (FFN) in each block.
The position-wise FFN consists of two fully connected layers with an activation function applied between them.
We use a Gaussian Error Linear Unit (GELU) activation~\cite{hendrycks2016bridging} instead of the standard ReLu activation, following the transformer in~\cite{radford2018improving}:
\begin{equation}
   \text{GELU}(x) = x\Phi(x), \quad \text{where} \ \Phi(x)=P(X\leq x), X \sim \mathcal{N}(0,1)
\end{equation}
The output of the FFN is a sequence of hidden states of the same length as the input:
\begin{equation}
    \text{FNN}(h^l_t) = \text{GELU}(h^l_tW^{(1)}+b^{(1)})W^{(2)}+b^{(2)}
\end{equation}
In addition, the element in the input sequence is transformed independently at each position, without considering the other elements in the sequence.

\subsection{Causal Layer}
The ultimate output, represented as $H^L_t = (G^L_{t-K+1}, s^L_{t-K+1}, a^L_{t-K+1}\\,..., G^L_t, s^L_t, a^L_t)$, is the final representation of the input trajectory after hierarchically processing it through all $L$ transformer blocks.
We segment the output into three sets: a set of RTGs, a set of states and a set of actions, which can be represented as:
\begin{equation}
    H^L_t = (G_{t-K+1:t}, s_{t-K+1:t}, a_{t-K+1:t})
\end{equation} 

These sets are further used to build two separate final representations for the action and state prediction.
As the action selection is based on the equation $a_{t} = \pi(s_{t-K+1:t}, G_{t-K+1:t})$, we use the hidden states $G_{t-K+1:t}$ and $s_{t-K+1:t}$ to build the final representations for the action prediction. The final representation is constructed by taking an element-wise summation of these two hidden states. A fully-connected linear layer and GELU activation are applied to obtain this representation:

\begin{equation}
    \begin{aligned}
    \tilde{\Psi}^a_t = G_{t-K+1:t} \oplus s_{t-K+1:t}\\
    \Psi^a_t = \text{GELU}(\tilde{\Psi}^a_t W^{r_1}+b^{r_1})
\end{aligned}
\end{equation}

Similarly, the final representations for the state prediction are constructed by using $s_{t-K+1:t}$ and $a_{t-K+1:t}$, because the transition of the state is based on its previous state and action:

\begin{equation}
    \begin{aligned}
    \tilde{\Psi}^s_t = s_{t-K+1:t} \oplus a_{t-K+1:t}\\
    \Psi^s_t = \text{GELU}(\tilde{\Psi}^a_t W^{r_2}+b^{r_2})
\end{aligned}
\end{equation}
Dropout~\cite{srivastava2014dropout} is also applied to the output of the fully-connected linear layer to help prevent overfitting.
Two fully-connected networks are put on top of the action prediction and state prediction for the final prediction task: predict the potential reward $r_t$ and generate action $a_t$.
The action prediction and state prediction are passed through the fully-connected networks, called reward estimation network $N_e$ to make the final predictions of potential reward $r_t$. And the potential reward $r_t$ and action prediction are received by the action generation network $N_g$ to generate action $a_t$.


\subsection{Training Procedure}
To train the model, we use a dataset of recommendation trajectories.
We first use Deep Deterministic Policy Gradient (DDPG)~\cite{lillicrap2015continuous} to train an expert RL agent, and then we use this expert in the environment to collect a series of expert trajectories to be the dataset.
We sample mini-batches of sequences with context length K from the dataset and reformulate the trajectories as the inputs of the transformer.
We summarize the training procedures for
CDT4Rec in Algorithm 1.
Let $\theta_e$ and $\theta_g$ denote the trainable parameters for the reward estimation network $N_e$ and action generation network $N_g$, respectively. Further, let $\theta_s$ and $\theta_a$ denote all trainable parameters for predicting the state and action.
We fit the reward estimation network $N_e$, state prediction, and action prediction by minimizing the factual loss of the reward:
\begin{equation}
\begin{aligned}
    &\mathcal{L}_{N_e}(\theta_e, \theta_s, \theta_a)\\
    =&\mathbb{E}_{(G,s,a)\sim \tau}\Bigg[\frac{1}{K}\Sigma_{k=1}^K\bigg(r_k - N_e\big(s_p^{k-K+1:k}(\theta_s), a_p^{k-K+1:k}(\theta_a);\theta_e\big)\bigg)^2\Bigg]
\end{aligned}
\end{equation}
For the action generation network $N_g$, we fit the $N_g$ to generate the final action by minimizing the cross entropy loss:
\begin{equation}
\begin{aligned}
    &\mathcal{L}_{N_g}(\theta_g, \theta_e, \theta_a)\\
    =&\frac{1}{K}\mathbb{E}_{(G,s,a)\sim \tau}\Bigg[-\Sigma_{k=1}^K \log N_g\big(\hat{r}^{k-K+1:k}(\theta_e), a_p^{k-K+1:k}(\theta_a);\theta_g \big)\Bigg]
\end{aligned}
\end{equation}
The overall training objective for CDT4Rec is:
\begin{equation}
\begin{aligned}
    \argmin _{\theta_e, \theta_s, \theta_a}
    \mathcal{L}_{N_e}(\theta_e, \theta_s, \theta_a) 
    + \argmin _{\theta_g, \theta_e, \theta_a}
    \mathcal{L}_{N_g}(\theta_g, \theta_e, \theta_a)
\end{aligned}
\end{equation}

\begin{algorithm}
    \label{alg:train}
    \SetKwInOut{Input}{input}

    \Input{Offline data $\mathcal{D}$, context length $K$, batch size $B$, number of iterations $n_{it}$, model parameters $\theta_e$, $\theta_g$, $\theta_a$, $\theta_s$\;
    }
    
    \caption{CDT4Rec Training}
    Compute the trajectory sampling probability $p(\tau)=|\tau|/\sum_{\tau \in \mathcal{D}}|\tau|$\;
    \For{$i=1,...,n_{it}$}{
        Sample B trajectories out of $\mathcal{D}$\;
        \For{each sampled trajectory $\tau$}{
            Compute the RTG sequence according to~\Cref{eq:G}\;
            Sample a sub-trajectory $\tau_K$ of length $K$ from $\tau$\;
        }
        a\_pred = $\text{Transformer}_{\theta_e, \theta_g, \theta_a}(\tau_K)$\;
        r\_esti = $\text{Transformer}_{\theta_e, \theta_s, \theta_a}(\tau_K)$\;
        $\theta_e, \theta_g, \theta_a, \theta_s \leftarrow$ one gradient update using the sampled $\tau_K$.
    }
\end{algorithm}

\section{Experiments}
In this section, we report the outcomes of experiments that focus on
the following three main research questions:
\begin{itemize}
    \item \textbf{RQ1}: How does CDT4Rec compare with other traditional deep RL algorithms in online recommendation environments and offline dataset environments?
    \item \textbf{RQ2}: How does the number of trajectories affect the performance of CDT4Rec compared to other offline reinforcement learning approaches?
    \item \textbf{RQ3}: How do the hyper-parameters affect the performance in the \emph{online} simulation environment? 
\end{itemize}
We concentrate our hyper-parameters study on online simulation settings since they are more closely suited to real-world environments, whereas offline datasets are fixed and do not reflect users' dynamic interests.

\subsection{Datasets and Environments}
In this section, we compare our proposed algorithm, CDT4Rec, with other state-of-the-art algorithms on real-world datasets and within an online simulation environment. We implement our model in PyTorch and conduct all our experiments on a server with two Intel Xeon CPU E5-2697 v2 CPUs with 6 NVIDIA TITAN X Pascal GPUs, 2 NVIDIA TITAN RTX, and 768 GB memory.

Firstly, we introduce six public real-world representative datasets from different recommendation domains for offline experiments which vary significantly in sparsity.~\Cref{tab:stat} contains the overall statistics of the datasets.
\begin{itemize}
    \item Amazon CD\footnote{https://nijianmo.github.io/amazon/index.html}:  This is a collection of product review datasets crawled from Amazon.com by~\cite{ni2019justifying}. They separated the data into various datasets based on Amazon's top-level product categories. We use the ``CD'' category in this work.
    \item LibraryThing: \textit{LibraryThing}\footnote{https://www.librarything.com/} is an online service to help people catalog their books easily. It includes social relationships between users and is frequently used to study social recommendation algorithms.
    \item MovieLens: This is a popular benchmark dataset for recommender systems. In this work, we will use two different scales of the MovieLens datasets, MovieLens-1M\footnote{https://grouplens.org/datasets/movielens/1m/} and MovieLens-20M\footnote{https://grouplens.org/datasets/movielens/20m/}. 
    \item GoodReads: It is a dataset from the book review website \textit{GoodReads}\footnote{https://www.goodreads.com/} by~\cite{wan2018item}. It contains information on users' multiple interactions regarding items, including rating, review text, etc.
    \item Netflix: This is a well-known benchmark dataset from the Netflix Prize Challenge\footnote{https://www.kaggle.com/datasets/netflix-inc/netflix-prize-data}. It only contains rating information.
    \item Book-Crossing\footnote{http://www2.informatik.uni-freiburg.de/~cziegler/BX/}: This is another book related dataset proposed by~\cite{ziegler2005improving}. This dataset is similar to MovieLens which only contains rating information.
\end{itemize}
\begin{table}[ht]
    \centering
    \caption{Statistics of the datasets used in our offline experiments.}
    \begin{tabular}{c|c|c|c|c}
    \hline 
        Dataset & \# of Users & \# of Items & \# of Ratings & Density  \\\hline
         Amazon CD & 75,258 & 64,443 & 3,749,004 & 0.08\% \\
         LibraryThing & 73,882 & 337,561 & 979,053 & 0.004\% \\
         MovieLens-1M & 6040 & 3900 & 1,000,209 & 4.24 \%\\
         GoodReads & 808,749 &  1,561,465 & 225,394,930 & 0.02\%\\
         MovieLens-20M &  138,493  & 27,278 & 20,000,263 & 0.53\%\\ 
         Netflix & 480,189  & 17,770 & 100,498,277 & 1.18\%\\
    \hline 
    \end{tabular}
    \label{tab:stat}
\end{table}

In order to evaluate the performance of the proposed method, we need to transfer those offline datasets into simulation environments so that the reinforcement learning agent can interact. 
We convert those recorded data into an interactive environment, following the existing works~\cite{chen2020knowledge}. Specifically, we adopt LSTM as the state encoder for temporal information to ensure the temporal relation can be captured. 
We convert the feedback data into a binary format by categorizing ratings above 75\% of the maximum rating as positive feedback, which will provide a positive incentive signal for the agent. Ratings lower than 75\% are classified as negative feedback.
The evaluation process is the same as described in~\cite{zhao2018recommendations}.

In addition, we also conduct an experiment on a real online simulation platform to validate the proposed method. We use VirtualTB~\cite{shi2019virtual} as the major online platform in this work. VirtualTB mimics a real-world online retail environment for recommender systems. It is trained using hundreds of millions of genuine Taobao data points, one of China's largest online retail sites.
The VirtualTB simulator creates a ``live'' environment by generating customers and interactions, allowing the agent to be tested with virtual customers and the recommendation system.
It uses the GAN-for-Simulating Distribution (GAN-SD) technique with an extra distribution constraint to produce varied clients with static and dynamic features. The dynamic attributes represent changing interests throughout an interactive process. It also employs the Multi-agent Adversarial Imitation Learning (MAIL) technique to concurrently learn the customers' policies and the platform policy to provide the most realistic interactions possible. 


In terms of evaluation metrics, we use the click-through rate (CTR) for the online simulation platform as the CTR is one of the built-in evaluation metrics of the VirtualTB simulation environment. 
For offline dataset evaluation, we employ a variety of evaluation metrics, including recall, precision,  and normalized discounted cumulative gain (nDCG).

\subsection{Baselines}
Most of the existing works are evaluating their methods on offline datasets, and very few works provide a public online simulator evaluation. 
As there are two types of experiments, we provide two sets of 
baselines to be used for different experimental settings. Firstly, we will introduce the baselines for the online simulator, which are probably the most popular benchmarks in reinforcement learning:
\begin{itemize}
    \item \textbf{Deep Deterministic Policy Gradient (DDPG)}~\cite{lillicrap2015continuous} is an off-policy method for environments with continuous action spaces. DDPG employs a target policy network to compute an action that approximates maximization to deal with continuous action spaces.
    \item \textbf{Soft Actor Critic (SAC)}~\cite{haarnoja2018soft} is an off-policy maximum entropy Deep Reinforcement Learning approach that optimizes a stochastic policy. It employs the clipped double-Q method and entropy regularisation that trains the policy to maximize a trade-off between expected return and entropy.
    \item \textbf{Twin Delayed DDPG (TD3)}~\cite{fujimoto2018addressing} is an algorithm that improves baseline DDPG performance by incorporating three key tricks: learning two Q-functions instead of one, updating the policy less frequently, and adding noise to the target action.
    \item \textbf{Decision Transformer (DT)}~\cite{chen2021decision} is an offline reinforcement learning algorithm that incorporates the transformer as the major network component to infer actions.
\end{itemize}
Moreover, the following recommendation algorithms are used for offline evaluations which come from two different categories: transformer-based methods and reinforcement learning-based methods.
\begin{itemize}
    \item \textbf{SASRec}~\cite{kang2018self} is a well-known baseline that uses the self-attention mechanism to make sequential recommendations.
    \item \textbf{BERT4Rec}~\cite{sun2019bert4rec} is a recent transformer based method for recommendation. It adopts BERT to build a recommender system.
    \item \textbf{S3Rec}~\cite{zhou2020s3} is BERT4Rec follow-up work that uses transformer architecture and self-supervised learning to maximize mutual information.
    \item \textbf{KGRL}~\cite{chen2020knowledge} is a reinforcement learning-based method that utilizes the capability of Graph Convolutional Network (GCN) to process the knowledge graph information.
    \item \textbf{TPGR}~\cite{chen2019large} is a model that uses reinforcement learning and binary tree for large-scale interactive recommendations.
    \item \textbf{PGPR}~\cite{xian2019reinforcement} is a knowledge-aware model that employs reinforcement learning for explainable recommendations.
\end{itemize}
We note that SASRec, BERT4Rec, and S3Rec are not suitable for the reinforcement learning evaluation procedure.
In order to evaluate the performance of those models, we feed the trajectory representation $\tau$ as an embedding into those models for training purposes and use the remaining trajectories for testing purposes
\footnote{\url{https://drive.google.com/file/d/1zFCWn_QPpqRGG1mxGt_YVYUcQmZn_3qE/view?usp=sharing}}.

\subsection{Online Simulator Experiments (RQ1)}
Firstly, we will outline the procedures for conducting an online simulator experiment using offline reinforcement learning-based methods. Unlike traditional reinforcement learning algorithms (DDPG, SAC, etc.), we begin by training an expert agent using DDPG. We then employ this expert in the environment to collect a set of expert trajectories. It is important to note that the expert can only access a portion of the environment since we are collecting a fixed number of random trajectories as recorded data. These expert trajectories are treated as background knowledge and used to pre-train the offline reinforcement learning-based methods. Subsequently, the offline reinforcement learning method will conduct fine-tuning during the interaction with the simulation environment.

\begin{figure}[!h]
    \centering
    \includegraphics[width=\linewidth]{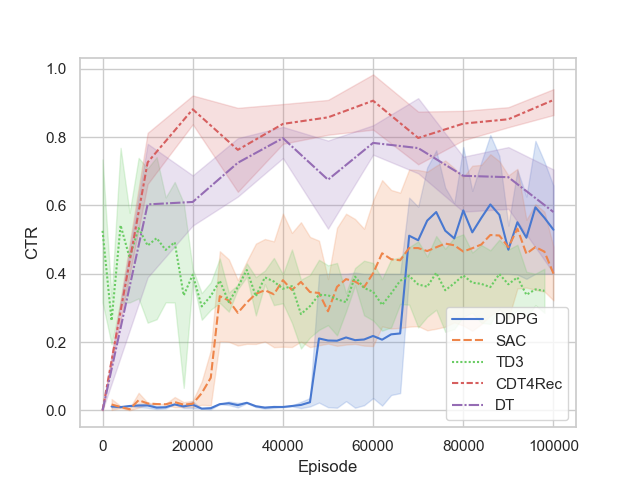}
    \caption{Overall comparison result with variance between the baselines and CDT4Rec in the VirtualTaobao simulation environment.}
    \label{fig:over_comp}
\end{figure}

In our online simulator experiments, we compared CDT4Rec with the aforementioned related reinforcement learning algorithms. The results of this comparison can be found in~\Cref{fig:over_comp}. We can clearly find that the proposed CDT4Rec method and DT lead to a significant improvement at the very start of the episode. 
One main reason is that DT and CDT4Rec can access more recorded offline trajectories from the simulation environments due to the offline RL configuration.

DT performs worse than CDT4Rec. One main reason is that CDT4Rec introduces an extra causal layer to estimate the potential causal effect on the users' decision process. Moreover, the introduction of cross-attention has apparently helped to generate better embeddings.

\subsection{Offline Dataset Experiments (RQ1)}
The overall results of the offline dataset experiments can be found in~\Cref{tab:result}. We can find that CDT4Rec outperforms all baselines, including transformer-based methods and reinforcement learning-based methods. We can see that on some datasets, transformer-based methods are better than reinforcement learning-based methods, but not to a significant extent. 
\begin{table*}[!ht]
\caption{The overall results of our model comparison with several state-of-the-art models on different datasets. The highest results are in bold and the second highest are marked by *}\smallskip
\begin{minipage}[ht]{1.0\linewidth}
\resizebox{\columnwidth}{!}{%
\begin{tabular}{ccccccc}
\hline
\multicolumn{1}{c|}{Dataset} & \multicolumn{3}{c|}{Amazon CD} & \multicolumn{3}{c}{Librarything} \\ \hline
\multicolumn{1}{c|}{Measure (\%)} & \multicolumn{1}{c|}{Recall} & \multicolumn{1}{c|}{Precision} & \multicolumn{1}{c|}{nDCG} & \multicolumn{1}{c|}{Recall} & \multicolumn{1}{c|}{Precision} & nDCG \\ \hline
\multicolumn{1}{c|}{SASRec} & 5.210 $\pm$ 0.202 & 2.352 $\pm$ 0.124 & \multicolumn{1}{c|}{4.601 $\pm$ 0.282 } & 8.312 $\pm$ 0.201 & 6.526 $\pm$ 0.129  & 7.391 $\pm$ 0.201  \\ 
\multicolumn{1}{c|}{BERT4Rec} & 9.123 $\pm$ 0.200 & 6.182 $\pm$ 0.211 & \multicolumn{1}{c|}{7.123 $\pm$ 0.198} & 11.982 $\pm$ 0.123 & 9.928 $\pm$ 0.201 & 10.021 $\pm$ 0.210\\ 
\multicolumn{1}{c|}{S3Rec} & 10.212 $\pm$ 0.192* & 7.928 $\pm$ 0.222* & \multicolumn{1}{c|}{8.028 $\pm$ 0.129*} & 13.425 $\pm$ 0.182 & 11.725 $\pm$ 0.182 & 11.237 $\pm$ 0.127  \\  
\multicolumn{1}{c|}{KGRL} & 8.208 $\pm$ 0.241 & 4.782 $\pm$ 0.341 & \multicolumn{1}{c|}{6.876 $\pm$ 0.511} & 12.128 $\pm$ 0.241 & 12.451 $\pm$ 0.242* & 13.925 $\pm$ 0.252* \\  
\multicolumn{1}{c|}{TPGR} & 7.294 $\pm$ 0.312 & 2.872 $\pm$ 0.531 & \multicolumn{1}{c|}{6.128 $\pm$ 0.541} & 14.713 $\pm$ 0.644* & 12.410 $\pm$ 0.612 & 13.225 $\pm$ 0.722 \\  
\multicolumn{1}{c|}{PGPR} & 6.619 $\pm$ 0.123 & 1.892 $\pm$ 0.143 & \multicolumn{1}{c|}{5.970 $\pm$ 0.131 } & 11.531 $\pm$ 0.241 & 10.333 $\pm$ 0.341 & 12.641 $\pm$ 0.442  \\  
\hline
\multicolumn{1}{c|}{CDT4Rec} & \textbf{10.424 $\pm$ 0.122} & \textbf{8.212 $\pm$ 0.201} & \multicolumn{1}{c|}{\textbf{8.111 $\pm$ 0.182}} & \textbf{15.229 $\pm$ 0.128} & \textbf{14.020 $\pm$ 0.201} & \textbf{14.768 $\pm$ 0.176}\\ 
\hline
\end{tabular}%
}
\end{minipage}

\begin{minipage}[ht]{1.0\linewidth}
\resizebox{\columnwidth}{!}{%
\begin{tabular}{ccccccc}
\hline
\multicolumn{1}{c|}{Dataset} &  \multicolumn{3}{c|}{Book-Crossing} & \multicolumn{3}{c}{GoodReads} \\ \hline
\multicolumn{1}{c|}{Measure (\%)} & \multicolumn{1}{c|}{Recall} & \multicolumn{1}{c|}{Precision} & \multicolumn{1}{c|}{nDCG} & \multicolumn{1}{c|}{Recall} & \multicolumn{1}{c|}{Precision} & nDCG \\ \hline
\multicolumn{1}{c|}{SASRec}& 5.831 $\pm$ 0.272  &  3.184 $\pm$ 0.149 &\multicolumn{1}{c|}{4.129 $\pm$ 0.390} & 6.921 $\pm$ 0.312 & 5.242 $\pm$ 0.211 & 6.124 $\pm$ 0.210  \\ 
\multicolumn{1}{c|}{BERT4Rec} & 8.222 $\pm$ 0.192 & 4.218 $\pm$ 0.129 & \multicolumn{1}{c|}{5.218 $\pm$ 0.129} & 8.483 $\pm$ 0.234 & 7.817 $\pm$ 0.281 & 8.012 $\pm$ 0.199  \\ 
\multicolumn{1}{c|}{S3Rec} & 8.992 $\pm$ 0.265* & 5.128 $\pm$ 0.239* & \multicolumn{1}{c|}{6.012 $\pm$ 0.200} & 10.263 $\pm$ 0.212 & 9.726 $\pm$ 0.188 & 10.002 $\pm$ 0.210*    \\  
\multicolumn{1}{c|}{KGRL} & 8.004 $\pm$ 0.223 & 3.521 $\pm$ 0.332 & \multicolumn{1}{c|}{7.641 $\pm$ 0.446*} & 7.459 $\pm$ 0.401 & 6.444 $\pm$ 0.321& 7.331 $\pm$ 0.301  \\  
\multicolumn{1}{c|}{TPGR} & 7.246 $\pm$ 0.321 & 4.523 $\pm$ 0.442 & \multicolumn{1}{c|}{6.870 $\pm$ 0.412} & 11.219 $\pm$ 0.323 & 10.322 $\pm$ 0.442* & 9.825 $\pm$ 0.642 \\  
\multicolumn{1}{c|}{PGPR} & 6.998 $\pm$ 0.112 & 3.932 $\pm$ 0.121 & \multicolumn{1}{c|}{6.333 $\pm$ 0.133} & 11.421 $\pm$ 0.223* & 10.042 $\pm$ 0.212 & 9.234 $\pm$ 0.242 \\  
\hline
\multicolumn{1}{c|}{CDT4Rec} & \textbf{9.234 $\pm$ 0.123} & \textbf{7.226 $\pm$ 0.289} & \multicolumn{1}{c|}{\textbf{8.276 $\pm$ 0.279}} & \textbf{13.274 $\pm$ 0.287} & \textbf{11.276 $\pm$ 0.175} & \textbf{10.768 $\pm$ 0.372}  \\ 
\hline
\end{tabular}%
}
\end{minipage}

\begin{minipage}[ht]{1.0\linewidth}
\resizebox{\columnwidth}{!}{%
\begin{tabular}{ccccccc}
\hline
\multicolumn{1}{c|}{Dataset} &  \multicolumn{3}{c|}{MovieLens-20M} & \multicolumn{3}{c}{Netflix} \\ \hline
\multicolumn{1}{c|}{Measure (\%)} & \multicolumn{1}{c|}{Recall} & \multicolumn{1}{c|}{Precision} & \multicolumn{1}{c|}{nDCG} & \multicolumn{1}{c|}{Recall} & \multicolumn{1}{c|}{Precision} & nDCG \\ \hline
\multicolumn{1}{c|}{SASRec} & 14.512 $\pm$ 0.510& 12.412 $\pm$ 0.333& \multicolumn{1}{c|}{12.401 $\pm$ 0.422} & 11.321 $\pm$ 0.231 & 10.322 $\pm$ 0.294   & 14.225 $\pm$ 0.421 \\ 
\multicolumn{1}{c|}{BERT4Rec}& 17.212 $\pm$ 0.233 & 14.234 $\pm$ 0.192 & \multicolumn{1}{c|}{13.292 $\pm$ 0.212} & 13.847 $\pm$ 0.128 & 12.098 $\pm$ 0.256 & 13.274 $\pm$ 0.210*  \\ 
\multicolumn{1}{c|}{S3Rec}& 17.423 $\pm$ 0.128* & 15.002 $\pm$ 0.221* & \multicolumn{1}{c|}{13.429 $\pm$ 0.520} & 14.090 $\pm$ 0.227* & 12.349 $\pm$ 0.256* & 13.002 $\pm$ 0.281  \\  
\multicolumn{1}{c|}{KGRL}  & 16.021 $\pm$ 0.498  & 14.989 $\pm$ 0.432 & \multicolumn{1}{c|}{13.007 $\pm$ 0.543} & 13.909 $\pm$ 0.343 & 11.874 $\pm$ 0.232 &13.082 $\pm$ 0.348 \\  
\multicolumn{1}{c|}{TPGR}  & 16.431 $\pm$ 0.369 & 13.421 $\pm$ 0.257 & \multicolumn{1}{c|}{13.512 $\pm$ 0.484*} & 12.512 $\pm$ 0.556 & 11.512 $\pm$ 0.595 & 10.425 $\pm$ 0.602 \\  
\multicolumn{1}{c|}{PGPR} & 14.234 $\pm$ 0.207 & 9.531 $\pm$ 0.219 & \multicolumn{1}{c|}{11.561 $\pm$ 0.228} & 10.982 $\pm$ 0.181 & 10.123 $\pm$ 0.227 & 10.104 $\pm$ 0.243 \\  
\hline
\multicolumn{1}{c|}{CDT4Rec} & \textbf{19.273 $\pm$ 0.212} & \textbf{17.371 $\pm$ 0.276} &  \multicolumn{1}{c|}{\textbf{17.311 $\pm$ 0.216}} & \textbf{15.271 $\pm$ 0.127} & \textbf{13.274 $\pm$ 0.168} & \textbf{12.479 $\pm$ 0.198}\\ 
\hline
\end{tabular}%
}
\end{minipage}
\label{tab:result}
\end{table*}
\subsection{Number of Trajectories Study (RQ2)}

\begin{figure*}[h]
     \centering
     \begin{subfigure}[b]{0.33\linewidth}
         \centering
         \includegraphics[width=\linewidth]{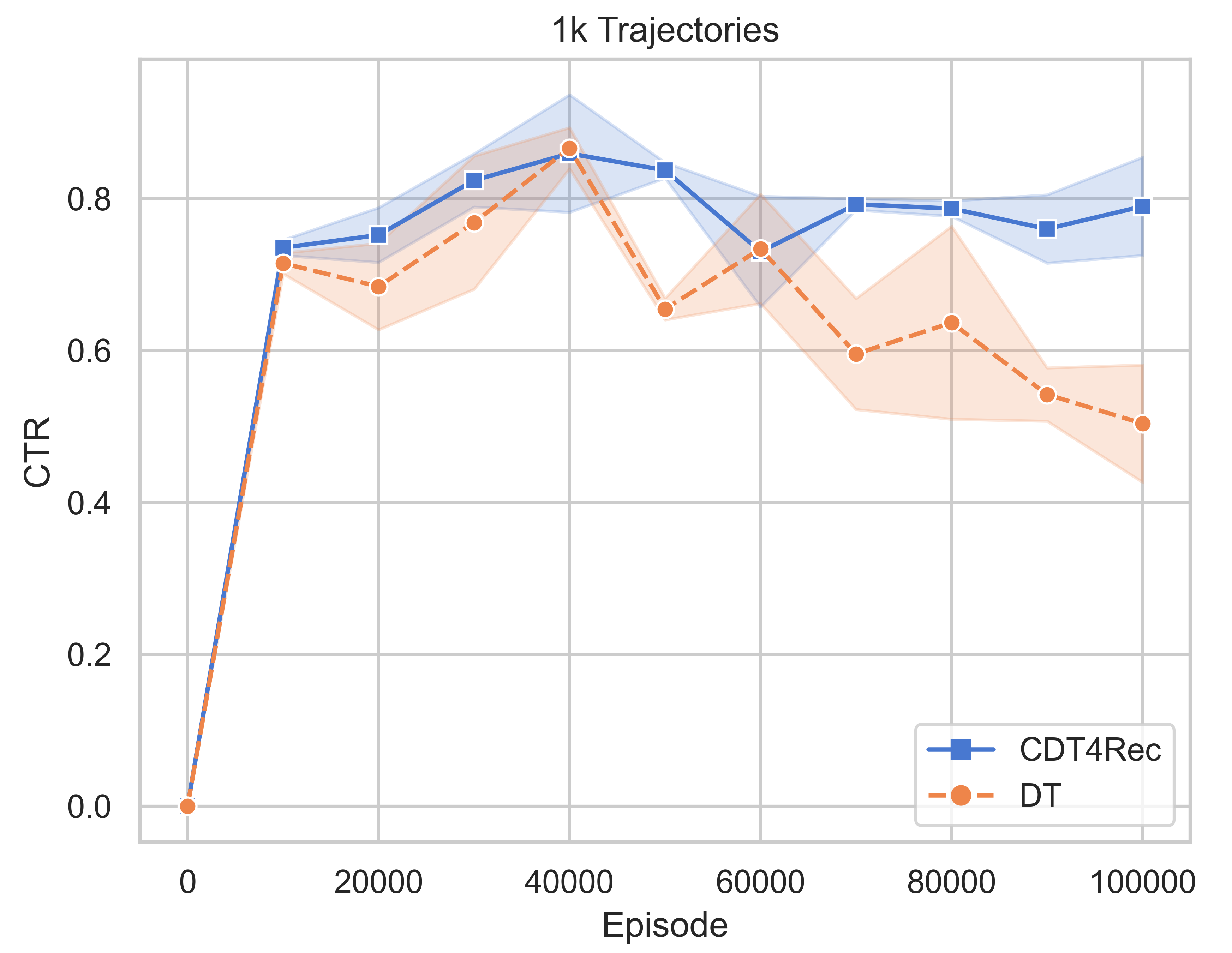}
     \end{subfigure}
     \begin{subfigure}[b]{0.33\linewidth}
         \centering
         \includegraphics[width=\linewidth]{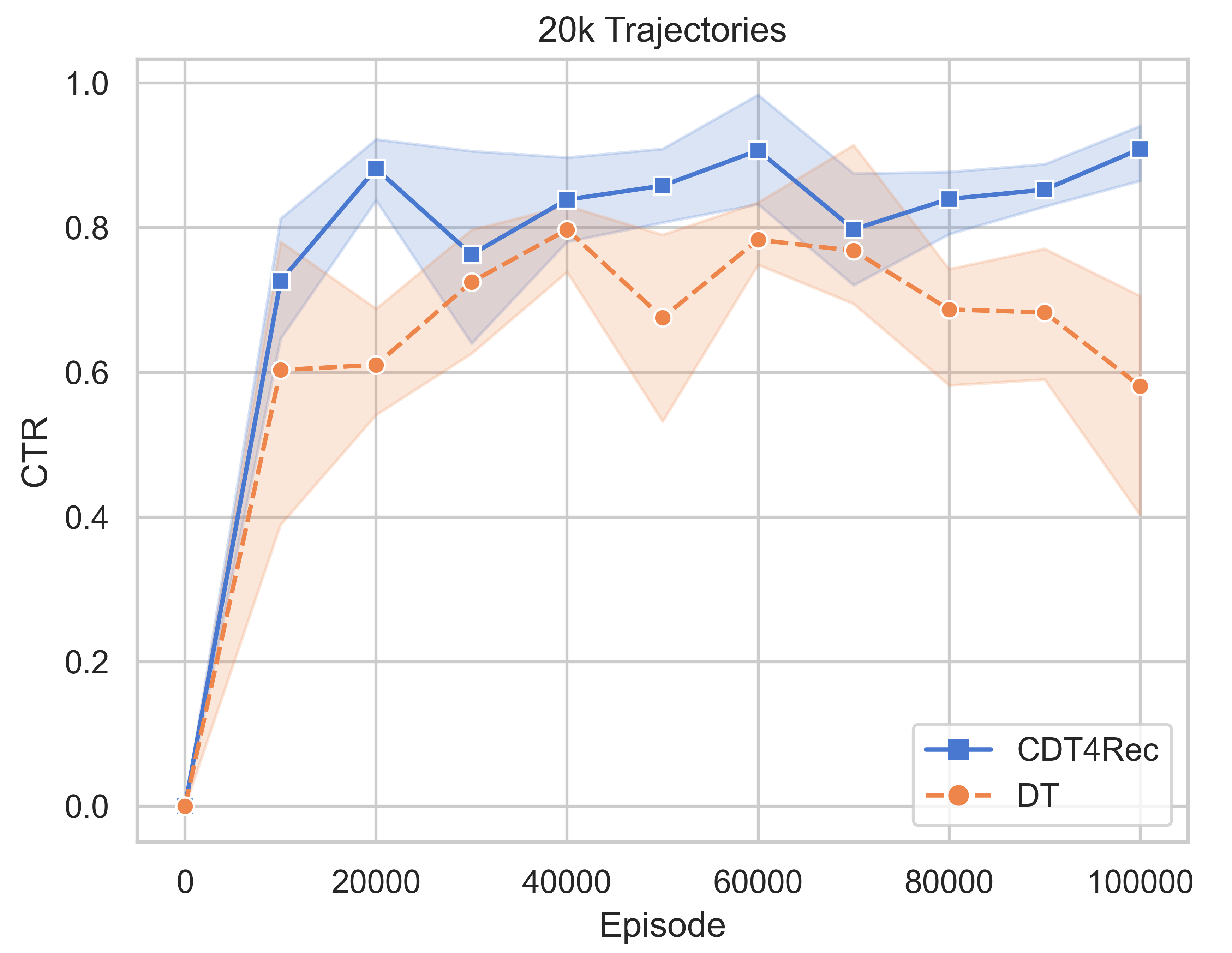}
     \end{subfigure}
     \begin{subfigure}[b]{0.33\linewidth}
         \centering
         \includegraphics[width=\linewidth]{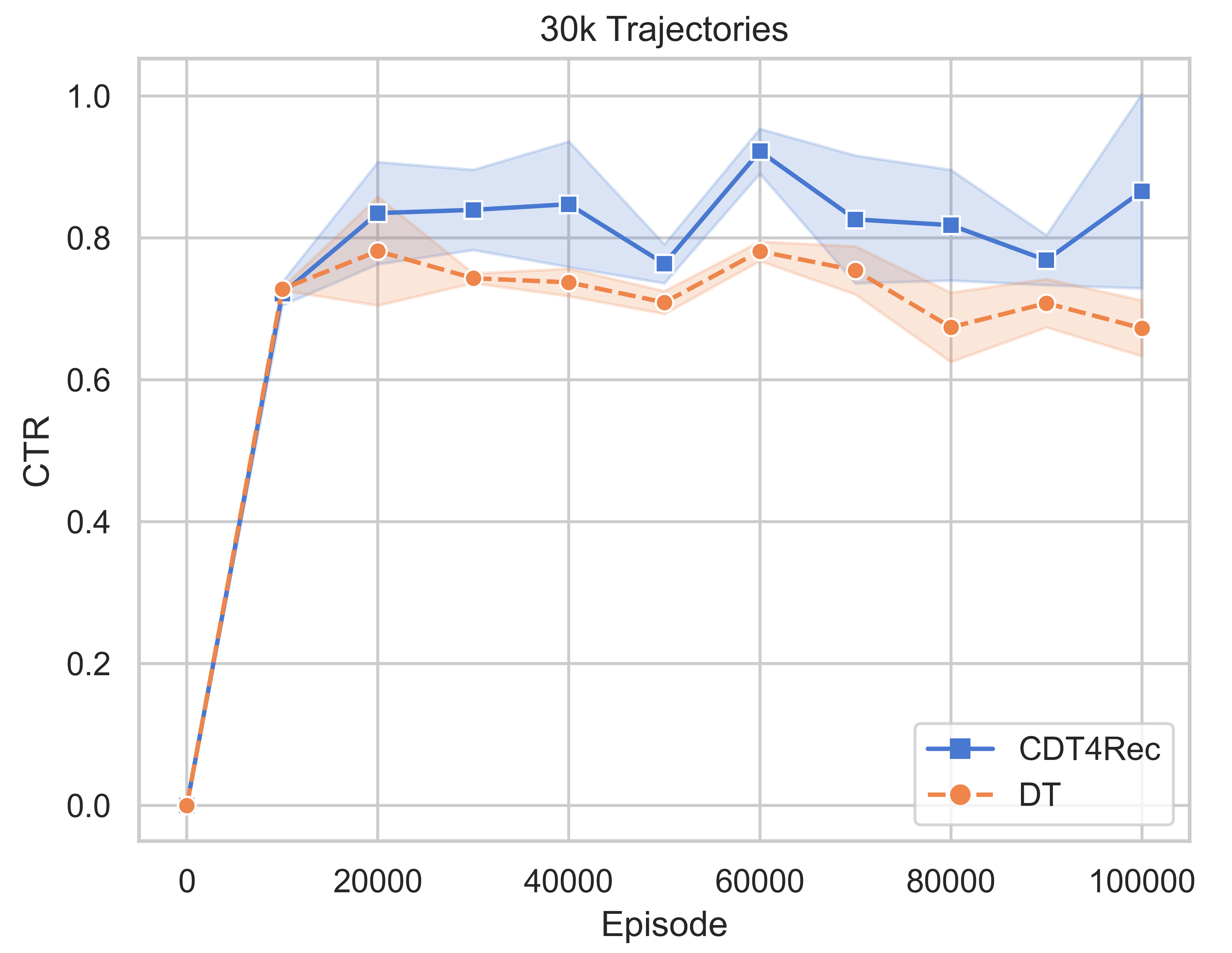}
     \end{subfigure}
        \caption{Performance Comparison Between CDT4Rec and DT for Different Numbers of Trajectories}
\label{fig:num_of_traje}
\end{figure*}

In order to answer this question (i.e., RQ2), we have conducted an experiment on different numbers of trajectories. We use DT as the major baseline, and the results are shown in~\Cref{fig:num_of_traje}. We can find that CDT4Rec outperforms DT over a range of trajectory lengths.
Moreover, when the number of trajectories is larger than 20k, the trend of CDT4Rec does not change significantly. We can conclude that CDT4Rec reaches its best performance with 20k trajectories.
Furthermore, we can see that the number of trajectories has no significant effect on CDT4Rec, as CDT4Rec performs well in all three numbers of trajectories,
indicating that CDT4Rec can learn effectively even with a smaller dataset.
While DT still has room for improvement, the performance of DT varies strongly, and the trend becomes more stable when the number of trajectories comes to 30k. 

\begin{figure*}[h]
     \begin{subfigure}[b]{0.38\linewidth}
         \centering         \includegraphics[width=\linewidth]{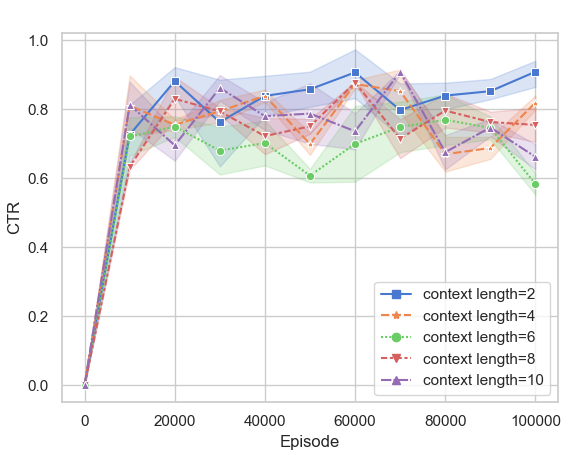}
         \caption{Context Length}
     \end{subfigure}
     \begin{subfigure}[b]{0.38\linewidth}
         \centering
         \includegraphics[width=\linewidth]{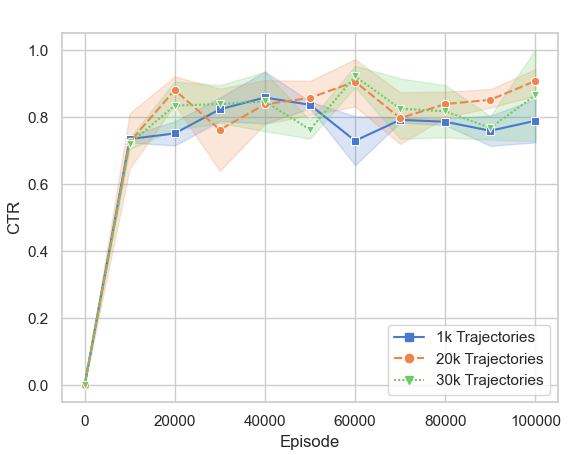}
         \caption{Number of Trajectories}
     \end{subfigure}
     \caption{Hyper-parameters Study for CDT4Rec}
     \label{fig:hyper}
\end{figure*}

\subsection{Hyper-parameters Study (RQ3)}
\label{sec:hyperpara}
In this section, we will investigate the impact of hyper-parameters. Here, we consider two different hyper-parameters: context length and the number of trajectories used for training. The context length determines the number of steps on which the agent's present action depends, and the number of trajectories means the number of trajectories included in the dataset. The results can be found in~\Cref{fig:hyper}. 
In our experiments, we find that when the context length is 2, CDT4Rec leads to the best performance. One plausible reason may be that user's behavior is highly determined by the most recent behavior (i.e., the previous one).  

Based on the results regarding the number of trajectories in the dataset, we observed that the performance did not vary considerably. This could be attributed to the fact that CDT4Rec is capable of learning effectively and performing well even with a smaller dataset.


\section{Related Work}
\vspace{1mm}\noindent\textbf{RL-based Recommender Systems.}
Reinforcement learning-based recommendation methods view the interactive process of making recommendations as a Markov Decision Process (MDP)~\cite{chen2021survey}. This approach can be divided into two categories: model-based and model-free methods.
\citet{bai2019model} proposed a model-based method that uses a generative adversarial training approach to jointly learn the user behavior model and update the policy for the recommendation.
Recently, there has been a trend in the literature toward using model-free techniques for RL-based recommendations.
\citet{zhao2018deep} 
proposed a page-wise recommendation framework based on deep RL to optimize the selection of items on a page by taking into account real-time feedback from users.
\citet{chen2020knowledge} introduced knowledge graphs into the RL framework to improve the efficiency of the decision-making process.
\citet{chen2021generative} designed a generative inverse reinforcement learning approach for online recommendation that automatically extracts a reward function from user behavior.
However, these are all online RL-based frameworks while we are interested in offline RL-based recommendations.

\vspace{1mm}\noindent\textbf{Offline RL.}
Recent studies have begun to investigate the possibility of integrating data-driven learning into the reinforcement learning framework.
\citet{kumar2020conservative} proposed conservative Q-learning for offline RL aiming to learn a lower bound of the value function.
MOReL~\cite{kidambi2020morel} is an algorithmic framework for model-based offline RL, which uses an offline dataset to learn a near-optimal policy by training on a Pessimistic MDP.
\citet{chen2021decision} presented an approach in which the offline RL problem is modeled as a conditional sequence modeling problem. They trained a transformer on collected offline data, utilizing a sequence modeling objective, with the goal of generating optimal actions.
\citet{janner2021offline} also adopted a transformer architecture and treat the offline RL as a sequence modeling problem.
\citet{kostrikov2021offline} introduced Fisher-BRC for offline RL that employs a simple critic representation and regularization technique to ensure that the learned policy remains consistent with the collected data.
However, these works are not applied for recommendation tasks and we are interested in the feasibility of employing offline RL for recommender systems.


\vspace{1mm}\noindent\textbf{Transformer in Recommender Systems.}
Recent developments in the field of sequential recommendations have seen a growing interest in incorporating transformer architectures.
\citet{sun2019bert4rec} proposed the model BERT4Rec that uses the bidirectional self-attention network to model user behavior sequences in sequential recommendation tasks.
\citet{wu2020sse} designed a personalized transformer architecture that effectively incorporates personalization in self-attentive neural network architectures, by incorporating SSE regularization.
\citet{chen2019behavior} introduced a method in which they incorporated the sequential signals of users' behavior sequences into a recommendation system on a real-world e-commerce platform. The method applied a self-attention mechanism to learn a better representation of each item in a user's behavior sequence by taking into account the sequential information.

\section{Conclusion}
In this work, we design a causal decision transformer for recommender systems (CDT4Rec) to address two major challenges: i) The difficulty of manually designing a reward function; ii) How to incorporate an offline dataset to reduce the amount of required expensive online interaction but maintain the performance. 
Our experimental results reveal that our proposed CDT4Rec model outperforms both existing RL algorithms and the state-of-the-art transformer-based
recommendation algorithms. 

In the future, we plan to investigate further how to better estimate  the causal effect of users' decisions so that we can better estimate the reward function via the collected trajectory. Moreover, we will work on further improving the offline reinforcement learning algorithm as we can find that there is still room for improvement.

\bibliographystyle{ACM-Reference-Format}
\bibliography{sample-base}
\end{document}